\documentclass[lineno]{biometrika}

\usepackage{pdfpages}
\AtEndDocument{
	\includepdf[pages=-]{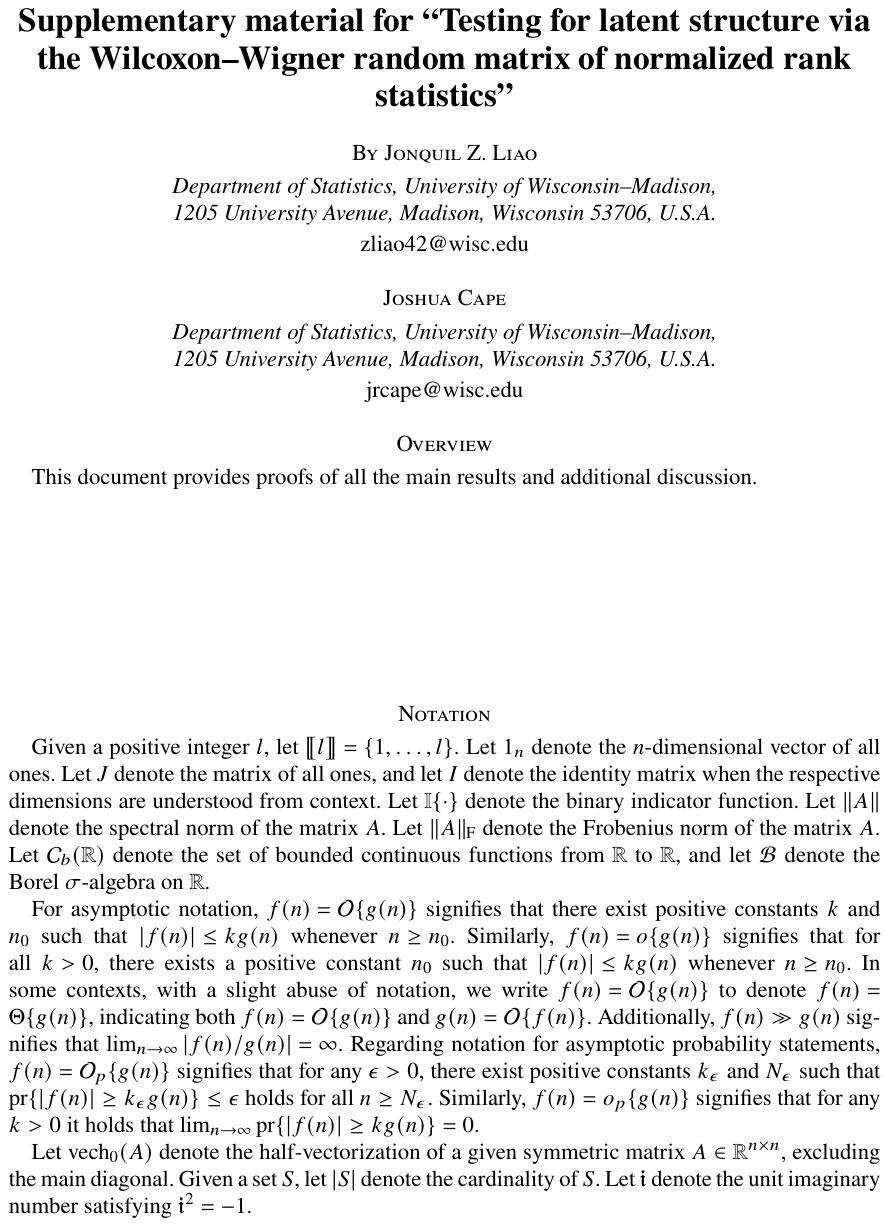}
}

\usepackage{amsmath}
\usepackage{newtxtext}
\usepackage[subscriptcorrection]{newtxmath}
\usepackage{xcolor}

\usepackage[colorlinks=true,citecolor=blue]{hyperref}%

\usepackage[capitalise]{cleveref}

\graphicspath{{./art/}}

\usepackage[plain,noend]{algorithm2e}

\makeatletter
\renewcommand{\algocf@captiontext}[2]{#1\algocf@typo. \AlCapFnt{}#2} % text of caption
\def\@algocf@capt@plain{top}
\renewcommand{\algocf@makecaption}[2]{%
  \addtolength{\hsize}{\algomargin}%
  \sbox\@tempboxa{\algocf@captiontext{#1}{#2}}%
  \ifdim\wd\@tempboxa >\hsize%     % if caption is longer than a line
    \hskip .5\algomargin%
    \parbox[t]{\hsize}{\algocf@captiontext{#1}{#2}}% then caption is not centered
  \else%
    \global\@minipagefalse%
    \hbox to\hsize{\box\@tempboxa}% else caption is centered
  \fi%
  \addtolength{\hsize}{-\algomargin}%
}
\makeatother

\def\T{{ \mathrm{\scriptscriptstyle T} }}

\def\pr{{\operatorname{pr}}}
\def\unif{{\operatorname{Unif}}}
\def\var{{\operatorname{var}}}
\def\cov{{\operatorname{cov}}}
\def\F{{\operatorname{F}}}

\begin{document}

\jname{Biometrika}
%% The year, volume, and number are determined on publication
\jyear{}
\jvol{}
\jnum{}
\cyear{}
%% The \doi{...} and \accessdate commands are used by the production team
%\doi{10.1093/biomet/asm023}
\accessdate{}

%% These dates are usually set by the production team
%\received{}
%\revised{}

%% The left and right page headers are defined here:
\markboth{Liao and Cape}{Testing for latent structure via the Wilcoxon--Wigner random matrix}

%% Here are the title, author names and addresses
\title{Testing for latent structure via the Wilcoxon--Wigner random matrix of normalized rank statistics}

\author{Jonquil Z. Liao}
\affil{Department of Statistics, University of Wisconsin--Madison,\\ 1205 University Avenue, Madison, Wisconsin 53706, U.S.A.
\email{zliao42@wisc.edu}}

\author{Joshua Cape}
\affil{Department of Statistics, University of Wisconsin--Madison,\\ 1205 University Avenue, Madison, Wisconsin 53706, U.S.A.
\email{jrcape@wisc.edu}}

\maketitle

\begin{abstract}
    This paper considers the problem of testing for latent structure in large symmetric data matrices. The goal here is to develop statistically principled methodology that is flexible in its applicability, computationally efficient, and insensitive to extreme data variation, thereby overcoming limitations facing existing approaches. To do so, we introduce and systematically study certain symmetric matrices, called Wilcoxon--Wigner random matrices, whose entries are normalized rank statistics derived from an underlying independent and identically distributed sample of absolutely continuous random variables. These matrices naturally arise as the matricization of one-sample problems in statistics and conceptually lie at the interface of nonparametrics, multivariate analysis, and data reduction. Among our results, we establish that the leading eigenvalue and corresponding eigenvector of Wilcoxon--Wigner random matrices admit asymptotically Gaussian fluctuations with explicit centering and scaling terms. These asymptotic results enable rigorous parameter-free and distribution-free spectral methodology for addressing two hypothesis testing problems, namely community detection and principal submatrix detection. Numerical examples illustrate the performance of the proposed approach. Throughout, our findings are juxtaposed with existing results based on the spectral properties of independent entry symmetric random matrices in signal-plus-noise data settings.
\end{abstract}


\section{Introduction}
\label{sec: intro}

\subsection{Overview}
Testing for latent structure in data matrices, often manifest via approximate low-rankness, has emerged as a common theme in various statistical problems of interest, including clustering, dimensionality reduction, and data denoising \citep{bouveyron2019model, wainwright2019high, chen2021spectral}.
Statistical network analysis is one particular research area with a longstanding interest in identifying and testing for the presence of latent structure, specifically by considering matrix-valued representations of networks.
There, the problem is traditionally formulated as distinguishing between a null hypothesis of no structure (i.e., homogeneity; a single community or block), typically represented by Erd\H{o}s--R\'{e}nyi random graphs \citep{erdhos1959random} or similar, versus an alternative hypothesis of particular structure (i.e., heterogeneity; multiple communities or blocks), typically represented by stochastic blockmodel random graphs or variants thereof \citep{holland1983stochastic}.

A focal aspect of this paper is to consider $\widetilde{R}$, a normalized rank-based transformation of an observable symmetric data matrix $A$, in lieu of $A$ itself.
Notably, it will be shown that the presence or absence of latent structure in $A$ can at times be inferred from the spectral properties or eigendecomposition of $\widetilde{R}$, even when latent structure is not readily discernible in the spectral properties of $A$ itself.
\cref{def: WW matrix} formally defines $\widetilde{R} \in [0,1]^{n \times n}$ which in the absence of underlying latent structure is herein named the \emph{Wilcoxon--Wigner random matrix of normalized rank statistics}. 
The proposed naming convention ``Wilcoxon--Wigner'' simultaneously acknowledges the contributions of Frank Wilcoxon to the development of rank-based tests in nonparametric statistics \citep{wilcoxon1945individual, wilcoxon1946group, wilcoxon1947probability} and of Eugene Wigner to the study of symmetric random matrices in physics and mathematics \citep{wigner1958distribution, wigner1967random}.

A key practical contribution of this paper is that the introduction of an ordinal, rank-based transformation in the formulation of Wilcoxon--Wigner random matrices leads to parameter-free, distribution-free nonparametric testing procedures, thereby circumventing the need to estimate nuisance parameters or make inflexible assumptions about data generating mechanisms.
In contrast, existing hypothesis testing procedures for random matrices and graphs often utilize estimated parameters to construct test statistics, such as in the stochastic blockmodel random graph framework where sample average-based probability and count estimates are employed for detecting the presence of a dense subgraph or determining the number of node communities \citep{lei2016goodness, bickel2016hypothesis, fan2022asymptotic, yuan2022hypothesis}.
Another key practical contribution is that our proposed low-rank spectral-based methodology is computationally efficient and transparent, avoiding the use of heuristic algorithms, permutation tests, or resampling procedures. 
In contrast, existing approaches for submatrix detection can be computationally intensive, time-consuming, or difficult to scale up to large datasets \citep{shabalin2009finding, butucea2013detection}.

Another methodological and practical advantage of using rank-based transformations is their robustness to extreme data variation.
When some or all entries of $A$ follow heavy-tailed distributions, conventional tests utilizing the spectrum of $A$ to detect low-rank structure or spikes in spiked models \citep{perry2018optimality, chung2019weak} are ineffective, as the spectra of matrices with heavy-tailed entry distributions exhibit significantly different behavior from those with light-tailed entries \citep{soshnikov2004poisson, auffinger2009poisson}.
In contrast, the rank transformation in \cref{def: WW matrix} does not require detailed prior knowledge about the population, is insensitive to heavy-tailed distributions, and promotes regularity which enables the study of matrix spectral properties via perturbation analysis.

\subsection{Context and content}
\label{sec: literature}
Recent years have witnessed flourishing research activity at the intersection of high-dimensional statistics and random matrix theory.
Spiked matrix models in high-dimensional settings, frequently conceptualized as population-level reference quantities corrupted by perturbations or noise, have attracted particular attention.
The most widely studied examples are spiked Wigner matrices and spiked covariance matrices.
These models are known to exhibit BBP-type (Baik--Ben Arous--P\'ech\'e) phase transition phenomena \citep{baik2005phase}.
Namely, the bulk spectrum asymptotically follows the semicircle or Marchenko--Pastur law for Wigner or Wishart random matrices, respectively, while spiked eigenvalues that exceed BBP thresholds separate from the bulk, and sub-critical spiked eigenvalues align at the edge of the bulk.
Additionally, the correlation between empirical eigenvectors and their theoretical counterparts becomes nontrivial for super-critical spikes, namely those that exceed certain known phase transition thresholds \citep{paul2007asymptotics, johnstone2018pca}.
These characteristics are extensively utilized throughout statistics in various detection, testing, and data denoising problems \citep{johnstone2018pca, perry2018optimality, bao2021singular}.
This paper shows that $\widetilde{R}$ can be viewed as a noisy approximately rank-one spiked matrix with diverging spike and further establishes the asymptotic distributional properties of the leading (i.e., largest) eigenvalue and corresponding eigenvector.

Rank statistics are widely encountered in the classical theory of robustness and nonparametrics.
The vast majority of rank-based tests involve statistics that are scalar-valued or vector-valued \citep{hajek1999theory}, with notable exceptions being Spearman's rho rank correlation matrix and Kendall's tau correlation matrix.
The spectral properties of these particular matrices have been extensively studied.
Namely, Spearman's rho rank correlation matrix asymptotically follows the Marchenko--Pastur law for its bulk spectrum \citep{bai2008large, wu2022limiting}, the Tracy--Widom law for its edge spectrum \citep{bao2019tracy}, a central limit theorem for its linear spectral statistics \citep{bao2015spectral}, and a Gumbel-type distribution for a polynomial of its largest off-diagonal entry \citep{zhou2007asymptotic}.
Similarly, Kendall's tau correlation matrix asymptotically follows an affine transformation of the Marchenko--Pastur law for its bulk spectrum \citep{bandeira2017marvcenko}, the Tracy--Widom law for its edge spectrum \citep{bao2019tracy2}, and a central limit theorem for its linear spectral statistics \citep{li2021central}.
The study of these matrices is driven in part by the need for robust nonparametric analogues of testing procedures, such as independence tests among the entries of random vectors using the spectral statistics of correlation matrices \citep{leung2018independence, li2021central}.
These matrices, while distinct from Wilcoxon--Wigner random matrices, similarly lie at the intersection of nonparametrics and high-dimensional statistics. 

This paper establishes that the spectral properties of Wilcoxon--Wigner random matrices can be leveraged to test hypotheses concerning latent population-level structure in two problems, namely community detection and principal submatrix detection.
In particular, the proposed eigenvalue-based test statistic is asymptotically standard normal under the null hypothesis of no latent structure and diverges under alternative hypotheses that are sufficiently well-separated from the null.
Here, as will be seen, alternative hypotheses correspond to the presence of unobserved block structure in the (consequently heterogeneous) observable data matrix $A$.
Notably, block structure can be viewed along the lines of a two-sample alternative hypothesis but with the crucial distinction that the labels are unknown to the analyst.

The results in this paper complement recent work on robust spectral clustering in \citet{cape2024robust}.
There, the authors similarly consider rank-transformed symmetric data matrices and perturbation analysis but for the different objective of (i) obtaining high-probability eigenvector-based misclustering error bounds, and (ii) describing the geometry of robust eigenvector-based data embeddings.
The contributions therein focus on estimation and clustering, whereas the results in the present paper concern hypothesis testing and inference.
In what follows, this distinction will be elaborated upon and made clear.

\subsection{Notation}
Given a positive integer $l$, let $\llbracket l \rrbracket = \{1, \ldots, l\}$.
Let $1_{n}$ denote the $n$-dimensional vector of all ones.
Let $J$ denote the matrix of all ones, and let $I$ denote the identity matrix when the respective dimensions are understood from context.
Let $\mathbb{I}\{\cdot\}$ denote the binary indicator function.
Let $\|A\|$ denote the spectral norm of the matrix $A$.
Let $\|A\|_{\F}$ denote the Frobenius norm of the matrix $A$.
Let $\mathcal{C}_{b}(\mathbb{R})$ denote the set of bounded continuous functions from $\mathbb{R}$ to $\mathbb{R}$, and let $\mathcal{B}$ denote the Borel $\sigma$-algebra on $\mathbb{R}$.

For asymptotic notation, $f(n) = \mathcal{O}\{g(n)\}$ signifies that there exist positive constants $k$ and $n_{0}$ such that $|f(n)| \leq k g(n)$ whenever $n \geq n_{0}$.
Similarly, $f(n) = o\{g(n)\}$ signifies that for all $k > 0$, there exists a positive constant $n_{0}$ such that $|f(n)| \leq k g(n)$ whenever $n \geq n_{0}$.
In some contexts, with a slight abuse of notation, we write $f(n) = \mathcal{O}\{g(n)\}$ to denote $f(n) = \Theta\{g(n)\}$, indicating both $f(n) = \mathcal{O}\{g(n)\}$ and $g(n) = \mathcal{O}\{f(n)\}$.
Additionally, $f(n) \gg g(n)$ signifies that $\lim_{n \to \infty}|f(n)/g(n)| = \infty$.
Regarding notation for asymptotic probability statements, $f(n) = \mathcal{O}_{p}\{g(n)\}$ signifies that for any $\epsilon > 0$, there exist positive constants $k_{\epsilon}$ and $N_{\epsilon}$ such that $\pr\{|f(n)| \geq k_{\epsilon} g(n)\} \leq \epsilon$ holds for all $n \geq N_{\epsilon}$.
Similarly, $f(n) = o_{p}\{g(n)\}$ signifies that for any $k > 0$ it holds that $\lim_{n \to \infty}\pr\{|f(n)| \geq k g(n)\} = 0$.

Proofs of the main results are provided in the Supplementary Material available online.
Code to reproduce the numerical examples is available online.

\section{Theory and methods}
\label{sec: theory methods}

\subsection{The Wilcoxon--Wigner random matrix of normalized rank statistics} 
\label{sec: WilcoxonWigner}
Let $\{A_{ij} : 1 \le i < j \le n\}$ denote a collection of $N = n(n-1)/2$ independent and identically distributed absolutely continuous random variables.
Herein, the \emph{Wilcoxon--Wigner random matrix of normalized rank statistics}, $\widetilde{R} \in [0,1]^{n \times n}$, is defined as the random matrix with discrete dependent entries given by
\begin{equation}\label{def: WW matrix}
    \widetilde{R}_{ij}
    =
    \begin{cases}
        (N+1)^{-1}\sum_{1 \le i^{\prime} < j^{\prime} \le n} \mathbb{I}\{A_{i^{\prime} j^{\prime}} \le A_{ij}\}
        &
        \text{if~~} i < j,\\
        \widetilde{R}_{ji}
        &
        \text{if~~} i > j,\\
        0
        &
        \text{if~~} i = j
        .
    \end{cases}
\end{equation}
In particular, for each $i < j$, the scaled entry $(N+1)\widetilde{R}_{ij} \in \{1, \ldots, N\}$ denotes the ordinal rank value of $A_{ij}$.

In this paper, the methodology developed around $\widetilde{R}$ is based on a systematic study of its spectral properties.
\cref{sec: esd and op} begins with a warm up by recording the entrywise properties of $\widetilde{R}$ and establishing that its whitened form matches the behavior of traditional Wigner random matrices, both asymptotically and non-asymptotically.
\cref{sec: normality} follows with a detailed investigation of the low-rank spectral properties of $\widetilde{R}$ (without whitening) which are subsequently leveraged for hypothesis testing.

\subsection{Preliminaries, empirical spectral distribution, and operator norm concentration}
\label{sec: esd and op}
For Wilcoxon--Wigner (WW) random matrices per \cref{sec: WilcoxonWigner}, each off-diagonal entry $\widetilde{R}_{ij}$ follows the discrete uniform distribution $\unif\{1/(N+1), \ldots, N/(N+1)\}$.
Consequently, direct computation reveals the following elementary properties.

\begin{proposition}[Entrywise properties for WW random matrices] \label{prop: entry property}
    Assume the setting in \cref{sec: WilcoxonWigner}.
    It holds that
    \begin{align*} 
        E(\widetilde{R}_{ij})
        &=
        1/2
        & \text{when~~} i \ne j, \\
        \var(\widetilde{R}_{ij})
        &=
        1/12 - 1/\{6(N+1)\}
        & \text{when~~} i \ne j,\\
        \cov(\widetilde{R}_{ij}, \widetilde{R}_{i'j'})
        &=
        -1/\{12(N+1)\}
        & \text{when~~} i \ne j, i' \ne j', \{i, j\} \ne \{i',j'\}
        .
    \end{align*}
\end{proposition}

Observe that the covariance appearing in \cref{prop: entry property} tends to zero as $n \to \infty$ while the variance is of constant order.
This weak dependence suggests the possibility that $\widetilde{R}$ might, after appropriate centering and scaling, exhibit global properties similar to independent entry Wigner ensembles.
\cref{thm: as convergence of operator norm} rigorously confirms this conjectured behavior.

\begin{theorem}[Semicircle law and Bai--Yin law for WW random matrices]
\label{thm: as convergence of operator norm}
    Assume the setting in \cref{sec: WilcoxonWigner}.
    Write $\sigma^{2}_{n} = \var(\widetilde{R}_{12})$, and define
    \begin{equation*}
        W
        \equiv
        W_{n}
        =
        \sigma_{n}^{-1}\{\widetilde{R} - E(\widetilde{R})\}
        .
    \end{equation*}
    In particular, $\var(W_{ij}) = 1$ for $i \ne j$.
    Let $\mu_{n}$ denote the empirical spectral distribution of $n^{-1/2}W$, and let $\mu$ denote the semicircle distribution \citep{wigner1958distribution} on $(\mathbb{R}, \mathcal{B})$.
    For all $f \in \mathcal{C}_{b}(\mathbb{R})$,
    \begin{equation*}
        \lim_{n \to \infty}
        \int_{\mathbb{R}} f(x) \mu_{n}(\mathrm{d}x)
        =
        \int_{\mathbb{R}} f(x) \mu(\mathrm{d}x)
        \qquad
        \text{almost surely}
        .
    \end{equation*}
    Furthermore,
    \begin{equation*}
        \lim_{n \to \infty}
        n^{-1/2}\| W \|
        =
        2
        \qquad
        \text{almost surely}
        .
    \end{equation*}
\end{theorem}

The upshot of \cref{thm: as convergence of operator norm} is that ranking-induced weak dependence here does not preclude the emergence of universality results known to hold for independent symmetric random matrix ensembles corresponding to mean-zero noise matrices in statistical models.
This section concludes with a non-asymptotic operator norm concentration inequality that further reinforces this point.

\begin{theorem}[Operator norm concentration for WW random matrices]\label{thm: operator norm concentration K1}
    Assume the setting in \cref{sec: WilcoxonWigner}.
    There exists a universal constant $C > 0$ such that for all $n \ge 2$,
    \begin{equation*}
        \pr
        \left(
        \| \widetilde{R} - E(\widetilde{R}) \|
        \geq
        6 n^{1/2}
        \right)
        \leq 
        \exp(-C n)
        .
    \end{equation*}
\end{theorem}

For context, \cref{thm: as convergence of operator norm,thm: operator norm concentration K1} are proved by showing that the macroscopic properties of $\widetilde{R}$ are well-approximated by those of a suitable Wigner-type proxy random matrix.
Such a proof strategy is adequate for \cref{sec: esd and op} but is no longer adequate when characterizing the distributional asymptotics of the leading eigenvalue and eigenvector in \cref{sec: normality}, where more direct, refined spectral perturbation analysis and consideration of entrywise dependence is required.

\subsection{Asymptotic normality of the leading eigenvalue and eigenvector}
\label{sec: normality}
By writing $\Gamma = \widetilde{R} - E(\widetilde{R})$, the Wilcoxon--Wigner random matrix $\widetilde{R}$ can be viewed as an additive perturbation of its expectation, i.e.,
\begin{equation}\label{eq: expect plus perturb}
    \widetilde{R}
    =
    E(\widetilde{R})
    +
    \Gamma
    ,
\end{equation}
where $E(\widetilde{R}) = 2^{-1}(J - I)$.
Equivalently, $\widetilde{R}$ may be described as a deterministic deformation of $\Gamma$ by $E(\widetilde{R})$.
In particular, Wilcoxon--Wigner random matrices resemble approximately rank-one spiked Wigner-type models with a diverging spike (i.e., leading eigenvalue) equal to $2^{-1}(n-1)$.

\cref{thm: largest eigenvalue} establishes that the leading eigenvalue of $\widetilde{R}$ admits asymptotically Gaussian fluctuations with closed-form expressions for the centering and scaling terms.

\begin{theorem}[Leading eigenvalue for WW random matrices]
\label{thm: largest eigenvalue} 
    Assume the setting in \cref{sec: WilcoxonWigner}.
    Let $\widehat{\lambda}_{1}(\widetilde{R})$ denote the leading eigenvalue of $\widetilde{R}$, and define $\widetilde{\sigma}_{n}^{2} = 8\sigma_{n}^{4}n^{-1}$, where $\sigma^{2}_{n} = \var(\widetilde{R}_{12})$.
    Then, as $n \to \infty$,
    \begin{equation*}
        \widetilde{\sigma}_{n}^{-1}
        \left\{
        \widehat{\lambda}_{1}(\widetilde{R}) - 2^{-1}(n-1) - 2\sigma_{n}^{2}
        \right\}
        \to
        N(0,1)
        \qquad
        \text{in distribution}
        .
    \end{equation*}
\end{theorem}

\cref{thm: largest eigenvalue} can be directly compared and contrasted to the corresponding classical asymptotic normality result for symmetric random matrices having independent, bounded entries.

\begin{theorem}[\citet{furedi1981eigenvalues}] \label{thm: FK}
    Let $A \in \mathbb{R}^{n \times n}$ be a symmetric random matrix whose entries $A_{ij}$ are bounded by a constant $K > 0$.
    Further, assume that
    $\{A_{ii} : 1 \leq i \leq n\}$ are independent and identically distributed with $E(A_{ii}) = v$, while
    $\{A_{ij} : 1 \leq i < j \leq n\}$ are independent and identically distributed with $E(A_{ij}) = \mu$ and $\var(A_{ij}) = \sigma^{2}$.
    Denote the eigenvalues of $A$ by $\widehat{\lambda}_{1}(A) \geq \cdots \geq \widehat{\lambda}_{n}(A)$.
    If $\mu > 0$, then as $n \to \infty$,
    \begin{equation*}
        2^{-1/2}\sigma^{-1}
        \left\{
        \widehat{\lambda}_{1}(A)
        -
        \mu(n-1)
        -
        v
        -
        \frac{\sigma^{2}}{\mu}
        \right\}
        \to
        N(0,1)
        \qquad
        \text{in distribution}
        .
    \end{equation*}
\end{theorem}

Observe that analogues of the entrywise properties $\mu$, $v$, and $\sigma^{2}$ for $A$ are given by $2^{-1}$, $0$, and $\sigma_{n}^{2}$ for $\widetilde{R}$.
As such, the centering terms agree for \cref{thm: largest eigenvalue,thm: FK}.
In contrast, the scaling terms differ for \cref{thm: largest eigenvalue,thm: FK}.
Namely, in \cref{thm: largest eigenvalue} the scaling is order $n^{1/2}$, whereas in \cref{thm: FK} the scaling is of constant order.
The underlying reason for this difference is briefly explained below in the following perturbation analysis that underlies the proofs.

For ease of discussion, let $\Gamma_{A} = A - E(A)$ and $\Gamma_{\widetilde{R}} = \widetilde{R} - E(\widetilde{R})$.
Let $\widehat{\lambda}_{1}(A)$ and $\widehat{\lambda}_{1}(\widetilde{R})$ denote the leading eigenvalues of $A$ and $\widetilde{R}$, respectively.
Let $\lambda_{1}\{E(A)\}$ and $\lambda_{1}\{E(\widetilde{R})\}$ denote the leading eigenvalues of the corresponding expectation matrices, respectively.
Write $u_{1} = n^{-1/2}1_{n}$.
For $A$ satisfying the hypotheses in \cref{thm: FK}, it holds that
\begin{equation*}
    \widehat{\lambda}_{1}(A) - \lambda_{1}\{E(A)\}
    =
    u_{1}^{\T} \Gamma_{A} u_{1}
    +
    \frac{u_{1}^{\T} \Gamma^{2}_{A} u_{1}}{\lambda_{1}\{E(A)\}}
    +
    \mathcal{O}_{p}(n^{-1})
    .
\end{equation*}
The leading order term $u_{1}^{\T} \Gamma_{A} u_{1}$ has variance $2\sigma^{2}$ as $n \to \infty$ which dominates the overall fluctuation.
In contrast, the property that $u_{1}^{\T} \Gamma_{\widetilde{R}} u_{1} = 0$ due to the dependence in $\widetilde{R}$ yields a different leading order term in the decomposition of $\widehat{\lambda}_{1}(\widetilde{R})$ given by
\begin{equation} \label{eq: decomp of lambda}
    \widehat{\lambda}_{1}(\widetilde{R}) - \lambda_{1}\{E(\widetilde{R})\}
    =
    \frac{u_{1}^{\T} \Gamma^{2}_{\widetilde{R}} u_{1}}{\lambda_{1}\{E(\widetilde{R})\}}
    +
    \mathcal{O}_{p}(n^{-1})
    .
\end{equation}
Here, the (vanishing) variance $8\sigma_{n}^{4}n^{-1}$ of $u_{1}^{\T} \Gamma^{2}_{\widetilde{R}} u_{1}/\lambda_{1}\{E(\widetilde{R})\}$ dictates the requisite scaling.

The decomposition in \cref{eq: decomp of lambda} is utilized throughout this paper to elucidate the limiting behavior of the leading eigenvalue as well as its corresponding eigenvector.
\cref{thm: largest eigenvector} quantifies the behavior of linear forms involving the leading eigenvector and establishes that they admit asymptotically Gaussian fluctuations.

\begin{theorem}[Linear forms of the leading eigenvector for WW random matrices]
\label{thm: largest eigenvector}
    Assume the setting in \cref{sec: WilcoxonWigner}.
    Let $\lambda_{1} = n/2$, let $u_{1} = n^{-1/2}1_{n}$, and let $\widehat{u}_{1}$ denote the leading unit norm eigenvector of $\widetilde{R}$ with choice of sign satisfying $u_{1}^{\T} \widehat{u}_{1} > 0$.
    Write $\Gamma \equiv \Gamma_{\widetilde{R}}$.
    If the deterministic unit vector $x$ satisfies $n \times \var(x^{\T}\Gamma u_{1}) \to \infty$, then
    \begin{equation*}
    \label{eq: vector xwu}
        x^{\T} \widehat{u}_{1}
        -
        \left\{1 - \frac{3 E(u_{1}^{\T} \Gamma^{2} u_{1})}{2\lambda_{1}^{2}} \right\}
        x^{\T} u_{1}
        -
        \frac{E(x^{\T} \Gamma^{2} u_{1})}{\lambda^{2}_{1}}
        =
        \frac{x^{\T} \Gamma u_{1}}{\lambda_{1}}
        +
        o_{p}\left[\left\{ \var\left(\frac{x^{\T} \Gamma u_{1}}{\lambda_{1}}\right) \right\}^{1/2} \right]
        .
    \end{equation*}
    Moreover, as $n \to \infty$,
    \begin{equation*}
        \frac{x^{\T} \widehat{u}_{1}
        -
        \left\{1 - 3 E(u_{1}^{\T} \Gamma^{2} u_{1})/(2\lambda_{1}^{2})\right\} x^{\T} u_{1}
        -
        E(x^{\T} \Gamma^{2} u_{1})/\lambda^{2}_{1}}{\left\{ \var(x^{\T} \Gamma u_{1}/\lambda_{1}) \right\}^{1/2}}
        \to
        N(0,1)
        \qquad
        \text{in distribution}
        .
    \end{equation*}
    If $x = u_{1}$, then
    \begin{equation}\label{eq: decomp of uu}
        u_{1}^{\T} \widehat{u}_{1}
        =
        1
        -
        \frac{u_{1}^{\T} \Gamma^{2} u_{1}}{2 \lambda_{1}^{2}}
        +
        \mathcal{O}_{p}(n^{-2})
        .
    \end{equation}
    Moreover, as $n \to \infty$,
    \begin{equation}\label{eq: uu normal}
        n \widetilde{\sigma}_n^{-1}
        \left(
        u_{1}^{\T}\widehat{u}_{1} - 1 + \frac{1}{6n}
        \right)
        \to
        N(0,1)
        \qquad
        \text{in distribution}
        .
    \end{equation}
\end{theorem}

In \cref{thm: largest eigenvector}, when $x$ satisfies $n \times \var(x^{\T} \Gamma u_{1}) \to \infty$, the leading term in the decomposition of the linear form $x^{\T} \widehat{u}_{1}$ is $x^{\T} \Gamma u_{1}/\lambda_{1}$.
Consequently, the asymptotic Gaussianity is characterized by a scaling of $\{\var(x^{\T} \Gamma u_{1}/\lambda_{1})\}^{-1/2} = \mathcal{O}(n)$.
In contrast, the choice $x = u_{1}$ does not satisfy the aforementioned variance condition.
In this situation, the leading term in the decomposition of $u_{1}^{\T} \widehat{u}_{1}$ is $u_{1}^{\T} \Gamma^{2} u_{1} / (2 \lambda_{1}^2)$, resulting in a scaling of $n\widetilde{\sigma}_{n}^{-1} = \mathcal{O}(n^{3/2})$ needed for asymptotic normality.
Put differently, \cref{eq: uu normal} establishes that the cosine of the angle between the population and sample leading eigenvector approaches one, has a bias of order $\mathcal{O}(n^{-1})$, and has asymptotically Gaussian fluctuations under the scaling $\mathcal{O}(n^{3/2})$.

The varied behavior of $x^{\T} \widehat{u}_{1}$ for different choices of input $x$ is consistent with previous findings in independent entry random matrix models.
Namely, in \citet[Theorem 2]{fan2022asymptotic}, two different behaviors of linear forms are observed depending on the magnitude of $\var(x^{\T}\widehat{u}_{k})$ for $1 \le k \le K$, where $K$ denotes the rank of the population matrix.
There, for independent entry models, $\var(x^{\T}\widehat{u}_{k}) = \mathcal{O}(\lambda_{k}^{-2})$ under certain conditions, while $\var(u_{k}^{\T} \widehat{u}_{k}) = \mathcal{O}(\alpha_{n}^2\lambda_{k}^{-4})$ for $x = u_{k}$, where $\alpha_{n}$ is associated with the perturbation matrix $\Gamma_{A}$.
In the context of the WW random matrix, defining a similar quantity $\alpha_{n}$ yields $\alpha_{n} = \mathcal{O}(n^{1/2})$.
Notably, substituting this $\alpha_{n}$ and $\lambda_{1} = \mathcal{O}(n)$ into the aforementioned variance formulas yields variance magnitudes matching those in \cref{thm: largest eigenvector}, namely, $\var(x^{\T}\widehat{u}_{1}) = \mathcal{O}(n^{-2})$ and $\var(u_{1}^{\T}\widehat{u}_{1}) = \mathcal{O}(n^{-3})$.

\begin{remark}[Eigenvalue--eigenvector perturbation relationship]\label{remark: linear relationship}
    \cref{eq: decomp of lambda} and \cref{eq: decomp of uu} together reveal a direct relationship between $\widehat{\lambda}_{1}(\widetilde{R})$ and $u_{1}^{\T} \widehat{u}_{1}$, namely
    \begin{equation*}
        u_{1}^{\T} \widehat{u}_{1}
        =
        \frac{-1}{(n-1)}\widehat{\lambda}_{1}(\widetilde{R})
        +
        \frac{3}{2}
        +
        \mathcal{O}_{p}(n^{-2})
        .
    \end{equation*}
    This relationship, arising in the context of normalized rank statistics, does not hold for symmetric random matrices with independent upper triangular entries.
\end{remark}

\begin{remark}[Comparing eigenvector linear forms]\label{remark: similarity in linear form}
    Suppose the symmetric random matrix $A \in \mathbb{R}^{n \times n}$ has diagonal entries equal to zero and has i.i.d above-diagonal entries sampled from $\unif(0,1)$.
    Using previous notational convention, write $A = 2^{-1}(J-I) + \Gamma_{A}$, where the population matrix is the same as that of $\widetilde{R}$ in \cref{eq: expect plus perturb}.
    Here, the non-trivial entries of $\Gamma_{A}$ and $\Gamma_{\widetilde{R}}$ each have asymptotic variances equal to $1/12$.
    Let $\widehat{u}_{A}$ be the leading unit norm eigenvector of $A$ with choice of sign satisfying $u_{1}^{\T} \widehat{u}_{A} > 0$.
    Then, for $\lambda_{1}$ and $\widetilde{\sigma}_{n}$ as in \cref{thm: largest eigenvector}, it holds that
    \begin{equation}
        \label{eq: decomp of eigenvec uniform}
        u_{1}^{\T} \widehat{u}_{A}
        =
        1
        -
        \frac{u_{1}^{\T} \Gamma^{2}_{A} u_{1}}{2 \lambda_{1}^{2}}
        +
        \mathcal{O}_{p}(n^{-2})
        ,
    \end{equation}
    and as $n \to \infty$,
    \begin{equation}
        \label{eq: asymp of eigenvec uniform}
        n\widetilde{\sigma}_{n}^{-1}
        \left(
        u_{1}^{\T}\widehat{u}_{A} - 1 + \frac{1}{6n}
        \right)
        \to
        N(0,1)
        \qquad
        \text{in distribution}
        .
    \end{equation}
    The above asymptotics match those of $\widetilde{R}$ in \cref{eq: decomp of uu} and \cref{eq: uu normal}.
    This agreement is noteworthy because, while dependence affects the scaling of the limiting distribution of the eigenvalue as shown by contrasting \cref{thm: largest eigenvalue} and \cref{thm: FK}, the variability of the linear form of the eigenvector remains unaffected.
    This is due to the fact that $u_{1}^{\T} \Gamma_{A} u_{1}$ and $u_{1}^{\T} \Gamma_{\widetilde{R}} u_{1}$ do not affect the leading order behavior of $u_{1}^{\T} \widehat{u}_A$ and $u_{1}^{\T} \widehat{u}_{1}$, respectively.
\end{remark}

Building upon the result for $u_{1}^{\T} \widehat{u}_{1}$ in \cref{thm: largest eigenvector}, it is possible to further establish an asymptotic normal approximation for $\|\widehat{u}_{1}\widehat{u}_{1}^{\T}\ - u_{1}u_{1}^{\T}\|^{2}_{\F}$ as follows.

\begin{corollary}[Leading eigenvector subspace perturbation for WW random matrices]\label{coroll: eigenspace}
    Under the assumptions and notation in \cref{thm: largest eigenvector}, it holds that
    \begin{equation*} 
        \|\widehat{u}_{1}\widehat{u}_{1}^{\T} - u_{1}u_{1}^{\T}\|^{2}_{\F}
        =
        \frac{2u_{1}^{\T} \Gamma^{2} u_{1}}{\lambda_{1}^{2}}
        +
        \mathcal{O}_{p}(n^{-2})
        .
    \end{equation*}
    Moreover, as $n \to \infty$, 
    \begin{equation*}
        n\widetilde{\sigma}_{n}^{-1}
        \left(
        \|\widehat{u}_{1}\widehat{u}_{1}^{\T}
        -
        u_{1}u_{1}^{\T}\|^{2}_{\F}
        -
        \frac{2}{3n}
        \right) 
        \to
        N(0,1)
        \qquad
        \text{in distribution}
        .
    \end{equation*}
\end{corollary}

\begin{remark}[Comparison of subspace recovery, with versus without ranks]
    It is possible to further compare \cref{coroll: eigenspace} to properties of sample eigenvectors computed from symmetric random matrices with independent entries.
    Concretely, suppose $A \in \mathbb{R}^{n \times n}$ is a symmetric random matrix with zero diagonal and i.i.d Gaussian above-diagonal entries from $N(\mu, \sigma^{2})$ with $\mu \neq 0$.
    Here, the expected squared projection distance between the leading eigenspace of $A$ and $E(A)$ is written as $E(\|\widehat{u}_{A}\widehat{u}_{A}^{\T} - u_{1}u_{1}^{\T}\|^{2}_{\F})$.
    Let $\widetilde{R}$ denote the entrywise rank transformation of $A$, and note that $u_{1} u_{1}^{\T}$ corresponds to the leading eigenspace of both $E(A)$ and $E(\widetilde{R})$.
    As shown in the Supplementary Material, the ratio of expected squared distances satisfies
    \begin{equation}\label{eq: ratio of mean projection}
        \frac{E(\|\widehat{u}_{\widetilde{R}} \widehat{u}_{\widetilde{R}}^{\T}
        -
        u_{1} u_{1}^{\T}\|^{2}_{\F})}
        {E(\|\widehat{u}_{A} \widehat{u}_{A}^{\T}
        -
        u_{1} u_{1}^{\T}\|^{2}_{\F})}
        \to
        \frac{\mu^{2}}{3\sigma^{2}}
    \end{equation}
    as $n \to \infty$.
    In words, on the basis of this ratio criterion and at the granularity of limiting constants, if the signal-to-noise ratio $\mu^{2}/(3\sigma^{2})$ exceeds $1$, then the truncated rank-one eigendecomposition of the original matrix is preferable for estimating the population-level one-dimensional subspace, all else equal.
    Otherwise, if the signal-to-noise ratio is below $1$, then the low-rank truncation of the rank-transformed data is preferable for subspace recovery.
\end{remark}

\begin{remark}[Novelty and additional comparison to prior work]
    \label{remark: comparison to Cape et al. (2024)}
    The main results herein extend and go beyond those for the one-block setting (i.e., $K=1$) found in \citet{cape2024robust}.
    \cref{thm: as convergence of operator norm}, establishing convergence of the empirical spectral distribution and operator norm after whitening, is new and does not have an analogue in the previous paper.
    \cref{thm: operator norm concentration K1} substantially improves upon Lemma~3 in the previous paper to yield the conventional rate-optimal operator norm bound $\|\widetilde{R} - E(\widetilde{R})\| = \mathcal{O}_{p}(n^{1/2})$ for $K=1$, while the Supplementary Material provides the corresponding improvement to \citet[Lemma 4]{cape2024robust} in the general setting $K \ge 1$.
    This improvement is crucial for obtaining the hypotheses testing guarantees in \cref{sec: application}.
    \cref{thm: largest eigenvalue,thm: largest eigenvector}, which precisely describe the distributional asymptotics of the leading eigenvalue and eigenvector of Wilcoxon--Wigner random matrices, enable hypothesis testing and inference, unlike Theorem~14 in the previous paper which establishes the asymptotic normality of individual (and row vectors of) eigenvector components, primarily to describe the low-dimensional geometry of robust data embeddings.
    \cref{coroll: eigenspace} yields a much more refined treatment of Frobenius norm perturbations for $K=1$ compared to Theorem~7 in the previous paper, and it provides a theoretical guarantee complementing the empirical investigations in Section~6 therein.
    \cref{sec: application} develops rigorous spectral-based hypothesis testing capabilities which are unavailable in and beyond the scope of \citet{cape2024robust}.
\end{remark}

\section{Testing statistical hypotheses} 
\label{sec: application}

\subsection{Eigenvalue-based test statistic}
This section considers the problem of testing whether the data matrix $A$ exhibits population-level latent low-rank structure.
In particular, we consider testing the hypotheses
\begin{equation}
    \label{eq: H0H1 community detection}
    H_{0}:
    F_{1} = F_{2}
    \qquad
    \text{versus}
    \qquad
    H_{1}:
    F_{1} \ne F_{2}
    ,
\end{equation}
where $F_{1}$ and $F_{2}$ denote absolutely continuous distributions as discussed in each of our two problem settings below.
To do so, we leverage knowledge of the asymptotic distribution for the leading eigenvalue of the corresponding Wilcoxon--Wigner random matrix, $\widetilde{R}$, per \cref{thm: largest eigenvalue}.
Specifically, the following subsections consider the eigenvalue-based test statistic given by
\begin{equation*}
    T_{n}(\widetilde{R})
    =
    \widetilde{\sigma}_{n}^{-1}
    \left\{
    \widehat{\lambda}_{1}(\widetilde{R})
    -
    2^{-1}(n-1)
    -
    2\sigma_{n}^{2}
    \right\}
    .
\end{equation*}
If desired, the analogous eigenvector-based testing procedure can be derived per \cref{remark: linear relationship} and \cref{thm: largest eigenvector}.

\subsection{Setting 1: Community detection}
\label{sec: testing symmetry}
Let $F_{1}$ and $F_{2}$ denote two absolutely continuous distributions, and let $\theta \in \{-1,1\}^{n}$ be a vector satisfying $\sum_{i = 1}^{n} \theta_{i} = 0$.
Suppose that the data matrix $A \in \mathbb{R}^{n \times n}$ takes the form
\begin{equation}
    \label{def: A symmetric blocks}
    A_{ij}
    \overset{\operatorname{ind}}{\sim}
    \begin{cases}
        F_{1}
        &
        \text{if~~} \theta_{i}\theta_{j} = 1, \;
        i \le j,\\
        F_{2}
        &
        \text{if~~} \theta_{i}\theta_{j} = -1, \;
        i \le j,
    \end{cases}
    \qquad
    A_{ji}
    =
    A_{ij}
    ,
\end{equation}
though in what follows, the diagonal entries of $A$ are immaterial.
In the language of random graph models, the symmetric matrix $A$ per \cref{def: A symmetric blocks} has latent population-level community or block structure and can be viewed as the adjacency matrix of an undirected edge-weighted random graph.
Viewing the row and column indices of $A$ as node labels for an underlying graph, the entries of $\theta$ identify the block memberships or communities of the nodes, with an equal number of nodes in each of the two blocks.
In words, if the row and column indices of an entry in $A$ belong to the same block, then the entry is drawn from the distribution $F_{1}$.
Otherwise, the corresponding entry is drawn from the distribution $F_{2}$.

If $F_{1} = F_{2}$, then obtaining $\widetilde{R}$ from $A$ per \cref{def: WW matrix} yields precisely the Wilcoxon--Wigner random matrix studied throughout this paper, hence \cref{thm: largest eigenvalue} ensures that $T_{n}(\widetilde{R})$ has an asymptotically standard normal distribution under the null hypothesis of no community structure.
In what follows, define $E_{1} F_{2} = \int_{\mathbb{R}} F_{2}(x) \mathrm{d}F_{1}(x)$, which can also be expressed as $E_{1} F_{2} = \pr(X_{2} \le X_{1})$ in terms of independent random variables $X_{1} \sim F_{1}$ and $X_{2} \sim F_{2}$.
If $F_{1} \neq F_{2}$, then in contrast, $T_{n}(\widetilde{R})$ diverges under alternative hypotheses for which $|E_{1} F_{2} - 1/2|$ is large.
\cref{prop: symmetric block} formally states this behavior.

\begin{proposition}[Test statistic properties for community detection]
    \label{prop: symmetric block}
    Suppose $A$ satisfies \cref{def: A symmetric blocks}.
    Under $H_{0}$ in \cref{eq: H0H1 community detection}, as $n \to \infty$,
    \begin{equation*}
        T_{n}(\widetilde{R})
        \to
        N(0,1)
        \qquad
        \text{in distribution}
        .
    \end{equation*}
    Under a sequence of alternatives $H_{1}^{(n)} : E_{1}^{(n)} F_{2}^{(n)} = 1/2 + \epsilon_{n}$, if $|\epsilon_{n}| \gg n^{-1/4}$, then as $n \to \infty$,
    \begin{equation*}
        |T_{n}(\widetilde{R})|
        \to
        \infty
        \qquad
        \text{in probability}
        .
    \end{equation*}
\end{proposition}

More generally, the Supplementary Material demonstrates that under \cref{def: A symmetric blocks}, $T_{n}(\widetilde{R})$ is asymptotically standard normal under the less stringent hypothesis $E_{1}F_{2} = 1/2$.

\subsection{Setting 2: Principal submatrix detection}
\label{sec: testing submatrix}
Let $F_{1}$ and $F_{2}$ be two absolutely continuous distributions, and let $l \in \{0,1\}^{n}$ be a vector with $n_{1}$ entries equal to $1$ and all remaining entries equal to $0$.
Suppose the data matrix $A \in \mathbb R^{n \times n}$ takes the form
\begin{equation}
    \label{def: denser subgraph}
    A_{ij}
    \overset{\operatorname{ind}}{\sim}
    \begin{cases}
        F_{1}
        &
        \text{if~~} l_{i}l_{j} = 1, \;
        i \le j,\\
        F_{2}
        &
        \text{if~~} l_{i}l_{j} = 0, \;
        i \le j,
    \end{cases}
    \qquad
    A_{ji}
    =
    A_{ij}
    ,
\end{equation}
though in what follows, the diagonal entries of $A$ are immaterial.
Here, $A$ contains a principal submatrix whose entries follow $F_{1}$ whereas the remaining matrix entries follow $F_{2}$.
In the language of random graph models, this latent submatrix can be interpreted as representing an anomalous weighted subgraph with cardinality $n_{1}$.
Viewing the row and column indices of $A$ as node labels for an underlying graph, the entries of $l$ identify the block memberships of the nodes.
This section considers the setting $n_{1} = o(n)$ which has previously been investigated for (unweighted) stochastic blockmodel random graphs \citep{arias2014community,verzelen2015community,fan2022asymptotic}.
\cref{prop: denser subgraph} establishes the asymptotic behavior of $T_{n}(\widetilde{R})$ in the present setting.

\begin{proposition}[Test statistic properties for principal submatrix detection]
    \label{prop: denser subgraph}
    Suppose $A$ satisfies \cref{def: denser subgraph}.
    Under $H_{0}$ in \cref{eq: H0H1 community detection}, as $n \to \infty$,
    \begin{equation*}
        T_{n}(\widetilde{R})
        \to
        N(0,1)
        \qquad
        \text{in distribution}
        .
    \end{equation*}
    Under a sequence of alternatives $H_{1}^{(n)} : E_{1}^{(n)} F_{2}^{(n)} = 1/2 + \epsilon_{n}$, if $|\epsilon_{n}| \gg n/n_{1}^{3/2}$ and $n_{1} = o(n)$, then as $n \to \infty$,
    \begin{equation*}
        |T_{n}(\widetilde{R})|
        \to
        \infty
        \qquad
        \text{in probability}
        .
    \end{equation*}
\end{proposition}

\subsection{Advantages and limitations of testing using the Wilcoxon--Wigner random matrix}
\label{sec: advantages}
\cref{sec: intro} provides introductory discussion of the advantages afforded by normalized rank-based transformations when testing for latent structure.
This section elaborates on both advantages and limitations of the proposed approach.

For latent low-rank block-type matrix detection problems, conventional methods rely on the difference between the expectations of the entry distributions \citep{mossel2015reconstruction,abbe2018community}.
For example, in \cref{sec: testing symmetry,sec: testing submatrix}, if the distributions $F_{1}$ and $F_{2}$ have identical expectations, then $E(A)$ is a constant, single-block matrix, whence conventional tests fail to detect multiple blocks in $A$.
The rank statistic-based test using $\widetilde{R}$, however, correctly distinguishes between $F_{1}$ and $F_{2}$ provided the difference $|E_{1}F_{2} - 1/2|$ is sufficiently large.
For instance, for $F_{1} = N(1,2)$ and $F_{2} = \text{Exponential}(1)$, even though both distributions have expectation equal to $1$, the test can still consistently detect the block structure as $n \to \infty$ since $|E_{1}F_{2} - 1/2|$ is constant order, thereby satisfying the well-separated alternative hypothesis conditions in \cref{prop: symmetric block} and \cref{prop: denser subgraph}.

There is a sizable existing literature on general submatrix detection and localization in data matrices, addressing minimaxity, relative efficiency, and even rank-based methods \citep{shabalin2009finding,butucea2013detection,ma2015computational,cai2017computational,arias2017distribution,arias2018distribution}.
The majority of these works focus on models with independent additive Gaussian noise, enabling the study of optimality, broadly defined, but at the expense of flexibility and robustness.
In contrast, our proposed methodology is applicable more generally, with a focus on flexibility and robustness, rather than model-specific optimality.
In this way, our work differs from what is seemingly the more conventional paradigm in detection and estimation problems, as noted in \citet{ma2015computational}, where ``one first establishes a minimax lower bound for any test or estimator [subject to model specifications] and then constructs a specific procedure which attains the lower bound within a constant or logarithmic factor.''

Though different in its outlook and objectives, this paper bears resemblance to \citet{perry2018optimality} which, in the context of principal component analysis, investigates pre-transforming matrix entries for optimally detecting a planted spike in non-Gaussian Wigner-type ensembles.
That said, when viewing $A = \textsf{Signal} + \textsf{Noise}$, there are several key differences.
First, our rank-based entrywise transformation does not require knowledge of the noise distribution, whereas their entrywise transformation involves the probability density function of the noise which is typically unavailable in practice.
Second, our proposed methodology does not require any finite moment assumptions for the noise distribution, whereas \citet[Assumption 4.3]{perry2018optimality} requires the noise distribution to have at least ten finite moments as well as a non-vanishing three-times continuously differentiable density function.
As such, our generalist approach affords additional flexibility and robustness, without pursuing optimal detection for a particular model.

\section{Numerical examples}
\label{sec: simulation}

\subsection{Simulation illustrations of the main theorems}
\begin{figure}
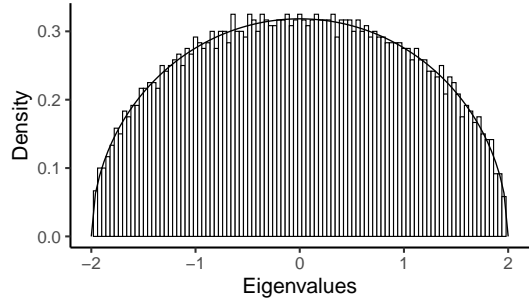

    \figuresize{.4}
    \figurebox{20pc}{25pc}{}[Wsemi]
    \caption{Empirical eigenvalue distribution for one realization of $n^{-1/2}W$ with $n = 3000$ compared to the semicircle distribution.}
    \label{fig1: histogram: semicircle}
\end{figure}

\begin{figure}
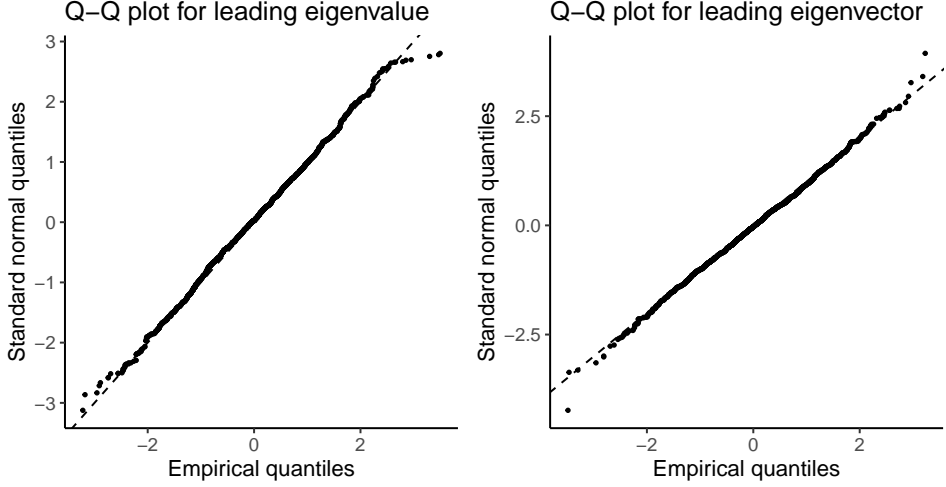

    \figuresize{.7}
    \figurebox{20pc}{25pc}{}[eigen]
    \caption{Left: Quantile--quantile plot for the standardized leading eigenvalue of $\widetilde{R}$ per \cref{thm: largest eigenvalue} compared to the standard normal distribution. Right: Quantile--quantile plot for the standardized leading eigenvector linear form per \cref{eq: uu normal} compared to the standard normal distribution. Here, $n = 2000$ with $2000$ replicates each.}
    \label{fig2: QQ eigenvalues and eigenvectors}
\end{figure}

This section presents numerical simulations that illustrate the main theorems in this paper.
\cref{fig1: histogram: semicircle} displays a normalized histogram of the eigenvalues for one realization of $n^{-1/2}W$ overlaid with the theoretical semicircle density shown by the solid curve.
\cref{fig2: QQ eigenvalues and eigenvectors} displays two quantile--quantile plots comparing the empirical and theoretical limiting distributions of the leading eigenvalue and eigenvector for WW random matrices, respectively.

\subsection{Variance transition from dependence to independence}
This subsection explores how the dependence among matrix entries influences the magnitude of the variance of the leading eigenvalue by constructing settings that interpolate between WW-type dependence and the independent entry regime.
As such, this subsection echoes the comparison of the scaling for the limiting distributions of $\widehat{\lambda}_{1}(\widetilde{R})$ and $\widehat{\lambda}_{1}(A)$ in \cref{sec: normality}.

For $\widetilde{R}$, the strictly upper triangular entries collectively constitute a uniform random permutation of $\{1/(N+1), \ldots, N/(N+1)\}$, and $\widetilde{\sigma}_{n}^{2} = \mathcal{O}(n^{-1})$ holds for $\widehat{\lambda}_{1}(\widetilde{R})$, per the discussion in \cref{sec: normality}.
In contrast, for a symmetric random matrix $U$ whose strictly upper triangular entries are i.i.d draws from $\unif(0,1)$, \cref{thm: FK} establishes that the variance of $\widehat{\lambda}_{1}(U)$ is of constant order $\mathcal{O}(1)$.
Here, we construct matrices $\widetilde{R}_{k} \in \mathbb{R}^{n \times n}$ that interpolate between these two regimes by manipulating the entrywise sampling scheme.
In particular, define $N_{k} = N + k$ and sample the $N$ strictly upper triangular entries of $\widetilde{R}_{k}$ from $\{1/(N_{k}+1), \ldots, N_{k}/(N_{k} + 1)\}$ without replacement.
Here, it is possible to show that the variance of $\widehat{\lambda}_{1}(\widetilde{R}_{k})$ satisfies
\begin{equation}\label{eq: variance k}
    \var\left\{\widehat{\lambda}_{1}(\widetilde{R}_{k})\right\}
    =
    \mathcal{O}
    \left[
    \max
    \left\{
    k(N+k)^{-1},
    n^{-1}
    \right\}
    \right]
    .
\end{equation}

In words, \cref{eq: variance k} shows a transition in the magnitude of the variance as $k$ grows from zero to infinity.
Observe that the matrix $\widetilde{R}_{k}$ is identically $\widetilde{R}$ when $k = 0$, whereas the entry distribution of $\widetilde{R}_{k}$ approaches that of $U$ as $k \to \infty$.
\cref{table: variance with different k} further illustrates these findings, where the estimated sample variances are calculated from $3000$ repeated simulations of the leading eigenvalue of $\widetilde{R}_{k}$ for different combinations of $k$ and $n$.

\begin{table}
    \tbl{Comparison of variance magnitude for $\widehat{\lambda}_{1}(\widetilde{R}_{k})$}
    {\begin{tabular}{ccccc}
    \hline
    \hline
    & $n = 1000$ & $n = 2000$ & $n = 4000$ & Magnitude of Variance \\ 
    \hline
    \hline
    $k = 0$ & 5.663e-05 & 2.804e-05 & 1.418e-05 & $\mathcal{O}(n^{-1})$\\
    $k = n$ & 3.871e-04 & 1.916e-04 & 9.781e-05 & $\mathcal{O}(n^{-1})$\\
    $k = n^{3/2}$ & 9.802e-03 & 7.333e-03 & 5.281e-03 & $\mathcal{O}(n^{-1/2})$\\
    $k = N$ & 8.470e-02 & 8.601e-02 & 8.112e-02 & $\mathcal{O}(1)$\\
    $k = \infty$ & 1.665e-01 & 1.650e-01 & 1.691e-01 & $\mathcal{O}(1)$\\
    \hline
    \hline
    \end{tabular}}
    \label{table: variance with different k}
    \begin{tabnote}
        Empirical variances of $\widehat{\lambda}_{1}(\widetilde{R}_{k})$ for different $k$ and $n$. The case $k = \infty$ corresponds to the matrix $U$. Calculations are based on $3000$ simulation replicates for each setting. 
    \end{tabnote}
\end{table}

\subsection{Simulation examples for community detection}
\label{sec: simulation symmetric block}

\begin{table}
    \tbl{Community detection using the leading eigenvalue statistic}
    {\begin{tabular}{ccccccc}
        \hline
        \hline
        & $\mu_{1} = \mu_{2}$ & $|E_{1}F_{2} - 1/2|$ & $n$ & $F_{1}$ & $F_{2}$ & Rejection Rate \\ 
        \hline
        \hline
        (a) & NA & $\mathcal{O}(1)$ & $2000$ & Pareto(1, 1) & $N(1, 0.1^{2})$ & 1 \\
        & NA & $\mathcal{O}(1)$ & $4000$ & Pareto(1, 1) & $N(1, 0.1^{2})$ & 1 \\
        (b) & Yes & $\mathcal{O}(1)$ & $2000$ & Pareto(1/2, 2) & $N(1, 0.1^{2})$ & 1 \\
        & Yes & $\mathcal{O}(1)$ & $4000$ & Pareto(1/2, 2) & $N(1, 0.1^{2})$ & 1 \\
        \hline
        (c) & No & $\mathcal{O}(1)$ & $2000$ & $N(1, 1)$ & $N(2, 1)$ & 1 \\
        & No & $\mathcal{O}(1)$ & $4000$ & $N(1, 1)$ & $N(2, 1)$ & 1 \\
        \hline
        (d) & No & $\mathcal{O}(n^{-1/8})$ & $2000$ & $N(1, 0.4^{2})$ & $N(1 + n^{-1/8}, 0.4^{2})$ & 1 \\
        & No & $\mathcal{O}(n^{-1/8})$ & $4000$ & $N(1, 0.4^{2})$ & $N( 1 + n^{-1/8}, 0.4^{2})$ & 1 \\
        (e) & No & $\mathcal{O}(n^{-1/4})$ & $2000$ & $N(1, 0.4^{2})$ & $N(1 + n^{-1/4}, 0.4^{2})$ & 0.180 \\
        & No & $\mathcal{O}(n^{-1/4})$ & $4000$ & $N(1, 0.4^{2})$ & $N(1 + n^{-1/4}, 0.4^{2})$ & 0.205 \\
        (f) & No & $\mathcal{O}(n^{-1/2})$ & $2000$ & $N(1, 0.4^{2})$ & $N(1 + n^{-1/2}, 0.4^{2})$ & 0.043 \\
        & No & $\mathcal{O}(n^{-1/2})$ & $4000$ & $N(1, 0.4^{2})$ & $N(1 + n^{-1/2}, 0.4^{2})$ & 0.065 \\
        \hline
        (g) & Yes & 0 & $2000$ & $N(1, 1)$ & $N(1, 2)$ & 0.040 \\
        & Yes & 0 & $4000$ & $N(1, 1)$ & $N(1, 2)$ & 0.053 \\
        \hline
        \hline
    \end{tabular}}
\label{table: application symmetric block}
\begin{tabnote}
    The left-most column indexes the experiments. Here, $\mu_{1}$ and $\mu_{2}$ denote the expectations of distributions $F_{1}$ and $F_{2}$, respectively. The second column indicates whether the two distributions share the same expectation, noting that expectations do not exist in Experiment (a). Rejection rates are calculated based on $400$ independent simulation replicates.
\end{tabnote}
\end{table}

We consider the asymptotically level $\alpha = 0.05$ two-tailed test that rejects $H_{0}$ when $|T_{n}(\widetilde{R})| > z_{0.025}$, where $z_{0.025}$ is the upper $0.025$ quantile of the standard normal distribution.
\cref{table: application symmetric block} displays the results of seven experiments with different choices of $F_{1}$ and $F_{2}$.
Each experiment is conducted with $n \in \{2000, 4000\}$, and the rejection rates are calculated based on $400$ independent simulation replicates.
These experiments convey several key messages.

The proposed testing procedure is consistent even in the presence of heavy tails.
In Experiment (a), the heavy-tailedness of $F_{1}$ results in the nonexistence of its expectation, and in Experiment (b), $F_{1}$ does not have finite variance.
Both experiments have empirical rejection rates of $1$, confirming that the proposed test consistently distinguishes between $F_{1}$ and $F_{2}$.

Experiment (c) illustrates our stated asymptotic guarantee in finite samples for two Gaussian distributions with well-separated expectations, though we emphasize that the proposed testing procedure does not rely on the difference of expectations.
In Experiments (b) and (g), the distributions $F_{1}$ and $F_{2}$ have the same expectation.
In the former setting, we see a rejection rate of $1$ due to the constant order gap $|E_{1}F_{2} - 1/2|$, whereas in the latter setting, we see a rejection rate near the nominal asymptotic level when $E_{1}F_{2} = 1/2$.

Experiments (d) through (f) manipulate the order of $|E_{1}F_{2} - 1/2|$ by controlling the expectation of $F_{2}$.
When $|E_{1}F_{2} - 1/2| \gg n^{-1/4}$, as in Experiment (d), the rejection rate is $1$, aligning with \cref{prop: symmetric block}.
In Experiments (e) and (f), where $|E_{1}F_{2} - 1/2| = \mathcal{O}(n^{-1/4})$, the rejection rate falls below $1$ but remains above or near the nominal asymptotic level.

\subsection{Simulation examples for principal submatrix detection}
\label{sec: simulation submatrix}

\begin{table}
    \tbl{Principal submatrix detection using the leading eigenvalue statistic}
    {\begin{tabular}{cccccccc}
    \hline
    \hline
    & $F_{1} = F_{2}$ & $|E_{1}F_{2} - 1/2|$ & $n$ & $n_{1}$ & $F_{1}$ & $F_{2}$ & Rejection Rate \\ 
    \hline  
    \hline
    (a) & Yes & $0$ & $2000$ & $300 \approx \mathcal{O}(n^{3/4})$ & Pareto(1, 1) & Pareto(1, 1) & 0.055 \\
    & Yes & $0$ & $4000$ & $500 \approx \mathcal{O}(n^{3/4})$ & Pareto(1, 1) & Pareto(1, 1) & 0.053 \\
    (b) & Yes & $0$ & $2000$ & $300 \approx \mathcal{O}(n^{3/4})$ & $N(1, 1)$ & $N(1, 1)$ & 0.065 \\
    & Yes & $0$ & $4000$ & $500 \approx \mathcal{O}(n^{3/4})$ & $N(1, 1)$ & $N(1, 1)$ & 0.055 \\
    \hline
    (c) & No & $\mathcal{O}(1)$ & $2000$ & $300 \approx \mathcal{O}(n^{3/4})$ & Pareto(1/2, 2) & $N(1, 1)$ & 0.995 \\
    & No & $\mathcal{O}(1)$ & $4000$ & $500 \approx \mathcal{O}(n^{3/4})$ & Pareto(1/2, 2) & $N(1, 1)$ & 1 \\
    (d) & No & $\mathcal{O}(1)$ & $2000$ & $300 \approx \mathcal{O}(n^{3/4})$ & Pareto(1, 1) & $N(1, 1)$ & 1\\
    & No & $\mathcal{O}(1)$ & $4000$ & $500 \approx \mathcal{O}(n^{3/4})$ & Pareto(1, 1) & $N(1, 1)$ & 1 \\
    \hline
    (e) & No & $\mathcal{O}(1)$ & $2000$ & $300 \approx \mathcal{O}(n^{3/4})$ & $N(2, 1)$ & $N(1, 1)$ & 1 \\
    & No & $\mathcal{O}(1)$ & $4000$ & $500 \approx \mathcal{O}(n^{3/4})$ & $N(2, 1)$ & $N(1, 1)$ & 1 \\
    (f) & No & $\mathcal{O}(1)$ & $2000$ & $40 \approx \mathcal{O}(n^{1/2})$ & $N(2, 1)$ & $N(1, 1)$ & 0.083 \\
    & No & $\mathcal{O}(1)$ & $4000$ & $60 \approx \mathcal{O}(n^{1/2})$ & $N(2, 1)$ & $N(1, 1)$ & 0.048 \\
    (g) & No & $\mathcal{O}(1)$ & $2000$ & $20 \approx \mathcal{O}(n^{2/5})$ & $N(2, 1)$ & $N(1, 1)$ & 0.040 \\
    & No & $\mathcal{O}(1)$ & $4000$ & $27 \approx \mathcal{O}(n^{2/5})$ & $N(2, 1)$ & $N(1, 1)$ & 0.033 \\
    \hline
    (h) & No & $\mathcal{O}(n^{-1/4})$ & $2000$ & $780 \approx \mathcal{O}(n^{7/8})$ & $N( 1 + n^{-1/4}, 1)$ & $N(1, 1)$ & 1 \\
    & No & $\mathcal{O}(n^{-1/4})$ & $4000$ & $1400 \approx \mathcal{O}(n^{7/8})$ & $N( 1 + n^{-1/4}, 1)$ & $N(1, 1)$ & 1 \\
    (i) & No & $\mathcal{O}(n^{-3/8})$ & $2000$ & $780 \approx \mathcal{O}(n^{7/8})$ & $N( 1 + n^{-3/8}, 1)$ & $N(1, 1)$ & 1 \\
    & No & $\mathcal{O}(n^{-3/8})$ & $4000$ & $1400 \approx \mathcal{O}(n^{7/8})$ & $N( 1 + n^{-3/8}, 1)$ & $N(1, 1)$ & 1 \\ 
    (j) & No & $\mathcal{O}(n^{-3/4})$ & $2000$ & $780 \approx \mathcal{O}(n^{7/8})$ & $N( 1 + n^{-3/4}, 1)$ & $N(1, 1)$ & 0.050 \\
    & No & $\mathcal{O}(n^{-3/4})$ & $4000$ & $1400 \approx \mathcal{O}(n^{7/8})$ & $N( 1 + n^{-3/4}, 1)$ & $N(1, 1)$ & 0.050 \\
    (k) & No & 0 & $2000$ & $780 \approx \mathcal{O}(n^{7/8})$ & $N(1, 1)$ & $N(1, 2)$ & 0.045 \\
    & No & 0 & $4000$ & $1400 \approx \mathcal{O}(n^{7/8})$ & $N(1, 1)$ & $N(1, 2)$ & 0.045 \\
    \hline
    \hline
\end{tabular}}
\label{table: application denser subgraph}
\begin{tabnote}
    The left-most column indexes the experiments. The row and column dimension of the submatrix is $n_{1}$, while $F_{1}$ denotes the corresponding entry distribution. Throughout, $n_{1} = o(n)$. Both $n_{1}$ and its magnitude relative to $n$ are presented. Rejection rates are calculated based on $400$ independent simulation replicates.
\end{tabnote}
\end{table}

We consider the asymptotically level $\alpha = 0.05$ two-tailed test that rejects $H_{0}$ when $|T_{n}(\widetilde{R})| > z_{0.025}$, where $z_{0.025}$ is the upper $0.025$ quantile of the standard normal distribution.
\cref{table: application denser subgraph} presents results for various experimental settings.
Both the exact value of $n_{1}$ and its magnitude relative to $n$ are displayed.
Throughout, $n_{1} = o(n)$ holds.
The experiments illustrate several important messages.

Experiments (a) and (b) serve as baselines, showcasing results when $F_{1}$ and $F_{2}$ are identical, irrespective of the moment properties of the data generating distributions.
The rejection rates in these experiments appear to stabilize around $0.05$ as $n$ increases, in accordance with the asymptotic distribution of the test statistic under $H_{0}$. 

Experiments (c) and (d) demonstrate that the test consistently detects the principal submatrix even when the distributions are heavy-tailed.
To reiterate, properties of the normalized rank statistics are insensitive to the tail properties of the original entry distributions.

Experiments (e) through (g) consider tests with the same set of underlying distributions while varying the magnitude of $n_{1}$ to corroborate \cref{prop: denser subgraph}.
Accordingly, when $|E_{1}F_{2} - 1/2|$ is of constant order and $n_{1}$ is order $n^{3/4}$, the condition $|E_{1}F_{2} - 1/2| \gg n/n_{1}^{3/2}$ is met, and a rejection rate of $1$ is anticipated and indeed observed in Experiment (e).
In Experiments (f) and (g), the order of $n_{1}$ is less than or equal to $n^{1/2}$, thus not meeting the aforementioned condition and explaining the low rejection rates.

Experiments (h) through (j) manipulate the magnitude of $|E_{1}F_{2} - 1/2|$ while keeping the magnitude of $n_{1}$ fixed.
Notably, Experiment (i) exhibits a rejection rate of $1$ even though the condition $|E_{1}F_{2} - 1/2| \gg n/n_{1}^{3/2}$ is not met.
This occurs because the (sufficient) gap condition in \cref{prop: denser subgraph} does not characterize the sharp detection boundary of the testing procedure.

\section{Discussion}
\label{sec: discussion}
The study of $\widetilde{R}$, with its dependent entries, forms the foundation of the proposed nonparametric latent structure testing methodology.
In recent years, interest has grown in the study of dependent random matrix ensembles in various settings and applications.
For example, \citet{anderson2008correlated} considers matrices with local dependence including so-called filtered Wigner matrices.
\citet{bryc2006hankel} studies matrices with Hankel, Markov, and Toeplitz patterned structure.
\citet{che2017universality} examines matrices with specific short-range correlation, while \citet{gotze2015limit, adamczak2011MPdependent} investigate matrices with martingale-type conditions imposed on their entries.
In a different context, \citet{agterberg2022entrywise} studies the entrywise perturbations of singular vectors when additive noise is permitted to exhibit heteroskedasticity and row-wise dependence.
For Wilcoxon--Wigner random matrices, the closest relevant setting of which we are aware is described in \citet{fleermann2019global, fleermann2021almost, fleermann2022central}, wherein weak forms of dependence are quantified via the separability of mixed moments.
Of course, $\widetilde{R}$ admits a particular form of structured dependence, while the more general treatment of random matrices with exchangeable entries is itself a separate topic of investigation \citep{chatterjee2006generalization}.

One distinguishing characteristic of $\widetilde{R}$ is that its spectral asymptotics are agnostic to the choice of absolutely continuous distribution for the entries of the data matrix $A$.
Here, continuity is assumed primarily for simplicity to avoid ties among $\{A_{ij} : 1 \le i < j \le n\}$.
In practice, if ties between data values are present then a random tie-breaking method could be applied.
For settings with numerous ties or tied observations, more careful considerations and domain-specific knowledge are needed, as is generally true for nonparametric rank-based tests \citep{savage1962bibliography}.

The introduction and analysis of (homogeneous) Wilcoxon--Wigner random matrices complements recent work by the authors that studies a robust spectral clustering method based on the rank-transform in \cref{def: WW matrix}.
Specifically, in the language of the present paper, \citet{cape2024robust} can be interpreted as studying heterogeneous variants of Wilcoxon--Wigner random matrices for which the set $\{A_{ij} : 1 \le i < j \le n\}$ is comprised of independent but not necessarily identically distributed elements.
The results in \citet{cape2024robust} are focused on quantifying the recovery of latent cluster structure at various levels of granularity.
In contrast, the present paper addresses hypothesis testing for community detection and principal submatrix detection via asymptotic theory under the baseline setting of no latent structure.
Taken together, these works shed new light on the intersection of spectral methods, perturbation analysis, and nonparametric statistics.

\section{Acknowledgement}
\label{sec: acknowledgement}
The authors thank the editor, associate editor, and reviewers for their detailed comments and helpful suggestions which improved this paper.
This work is supported in part by the National Science Foundation under grant DMS 2413552 and by the University of Wisconsin--Madison, Office of the Vice Chancellor for Research and Graduate Education, with funding from the Wisconsin Alumni Research Foundation.

\bibliographystyle{biometrika}
\bibliography{paper-ref}

\begin{thebibliography}{55}
\expandafter\ifx\csname natexlab\endcsname\relax\def\natexlab#1{#1}\fi

\bibitem[{Abbe(2018)}]{abbe2018community}
\textsc{Abbe, E.} (2018).
\newblock {Community detection and stochastic block models: recent
  developments}.
\newblock \textit{Journal of Machine Learning Research} \textbf{18}, 1--86.

\bibitem[{Adamczak(2011)}]{adamczak2011MPdependent}
\textsc{Adamczak, R.} (2011).
\newblock {On the Marchenko--Pastur and circular laws for some classes of
  random matrices with dependent entries}.
\newblock \textit{Electronic Journal of Probability} \textbf{16}, 1065--1095.

\bibitem[{Agterberg et~al.(2022)Agterberg, Lubberts \&
  Priebe}]{agterberg2022entrywise}
\textsc{Agterberg, J.}, \textsc{Lubberts, Z.} \& \textsc{Priebe, C.~E.} (2022).
\newblock Entrywise estimation of singular vectors of low-rank matrices with
  heteroskedasticity and dependence.
\newblock \textit{IEEE Transactions on Information Theory} \textbf{68},
  4618--4650.

\bibitem[{Anderson \& Zeitouni(2008)}]{anderson2008correlated}
\textsc{Anderson, G.} \& \textsc{Zeitouni, O.} (2008).
\newblock {A law of large numbers for finite-range dependent random matrices}.
\newblock \textit{Communications on Pure and Applied Mathematics} \textbf{61},
  1118--1154.

\bibitem[{Arias-Castro et~al.(2018)Arias-Castro, Castro, T{\'a}nczos \&
  Wang}]{arias2018distribution}
\textsc{Arias-Castro, E.}, \textsc{Castro, R.~M.}, \textsc{T{\'a}nczos, E.} \&
  \textsc{Wang, M.} (2018).
\newblock {Distribution-free detection of structured anomalies: permutation and
  rank-based scans}.
\newblock \textit{Journal of the American Statistical Association}
  \textbf{113}, 789--801.

\bibitem[{Arias-Castro \& Liu(2017)}]{arias2017distribution}
\textsc{Arias-Castro, E.} \& \textsc{Liu, Y.} (2017).
\newblock Distribution-free detection of a submatrix.
\newblock \textit{Journal of Multivariate Analysis} \textbf{156}, 29--38.

\bibitem[{Arias-Castro \& Verzelen(2014)}]{arias2014community}
\textsc{Arias-Castro, E.} \& \textsc{Verzelen, N.} (2014).
\newblock {Community detection in dense random networks}.
\newblock \textit{Annals of Statistics} \textbf{42}, 940--969.

\bibitem[{Auffinger et~al.(2009)Auffinger, Ben~Arous \&
  P\'ech\'e}]{auffinger2009poisson}
\textsc{Auffinger, A.}, \textsc{Ben~Arous, G.} \& \textsc{P\'ech\'e, S.}
  (2009).
\newblock {Poisson convergence for the largest eigenvalues of heavy-tailed
  random matrices}.
\newblock \textit{Annales de l’Institut Henri Poincaré - Probabilités et
  Statistique} \textbf{45}, 589--610.

\bibitem[{Bai \& Zhou(2008)}]{bai2008large}
\textsc{Bai, Z.} \& \textsc{Zhou, W.} (2008).
\newblock {Large sample covariance matrices without independence structures in
  columns}.
\newblock \textit{Statistica Sinica} \textbf{18}, 425--442.

\bibitem[{Baik et~al.(2005)Baik, Arous \& P{\'e}ch{\'e}}]{baik2005phase}
\textsc{Baik, J.}, \textsc{Arous, G.~B.} \& \textsc{P{\'e}ch{\'e}, S.} (2005).
\newblock {Phase transition of the largest eigenvalue for nonnull complex
  sample covariance matrices}.
\newblock \textit{Annals of Probability} \textbf{33}, 1643--1697.

\bibitem[{Bandeira et~al.(2017)Bandeira, Lodhia \&
  Rigollet}]{bandeira2017marvcenko}
\textsc{Bandeira, A.~S.}, \textsc{Lodhia, A.} \& \textsc{Rigollet, P.} (2017).
\newblock {Marcenko--Pastur law for Kendall's tau}.
\newblock \textit{Electronic Communications in Probability} \textbf{22}, 1--7.

\bibitem[{Bao(2019{\natexlab{a}})}]{bao2019tracy2}
\textsc{Bao, Z.} (2019{\natexlab{a}}).
\newblock {Tracy--Widom limit for Kendall’s tau}.
\newblock \textit{Annals of Statistics} \textbf{47}, 3504--3532.

\bibitem[{Bao(2019{\natexlab{b}})}]{bao2019tracy}
\textsc{Bao, Z.} (2019{\natexlab{b}}).
\newblock {Tracy--Widom limit for Spearman’s rho}.
\newblock \textit{Technical Report: see
  \text{https://sites.google.com/view/zhigangbaohomepage/}} .

\bibitem[{Bao et~al.(2021)Bao, Ding \& Wang}]{bao2021singular}
\textsc{Bao, Z.}, \textsc{Ding, X.} \& \textsc{Wang, K.} (2021).
\newblock {Singular vector and singular subspace distribution for the matrix
  denoising model}.
\newblock \textit{Annals of Statistics} \textbf{49}, 370--392.

\bibitem[{Bao et~al.(2015)Bao, Lin, Pan \& Zhou}]{bao2015spectral}
\textsc{Bao, Z.}, \textsc{Lin, L.-C.}, \textsc{Pan, G.} \& \textsc{Zhou, W.}
  (2015).
\newblock {Spectral statistics of large dimensional Spearman’s rank
  correlation matrix and its application}.
\newblock \textit{Annals of Statistics} \textbf{43}, 2588--2623.

\bibitem[{Bickel \& Sarkar(2016)}]{bickel2016hypothesis}
\textsc{Bickel, P.~J.} \& \textsc{Sarkar, P.} (2016).
\newblock {Hypothesis testing for automated community detection in networks}.
\newblock \textit{Journal of the Royal Statistical Society Series B:
  Statistical Methodology} \textbf{78}, 253--273.

\bibitem[{Bouveyron et~al.(2019)Bouveyron, Celeux, Murphy \&
  Raftery}]{bouveyron2019model}
\textsc{Bouveyron, C.}, \textsc{Celeux, G.}, \textsc{Murphy, T.~B.} \&
  \textsc{Raftery, A.~E.} (2019).
\newblock \textit{{Model-based clustering and classification for data science
  with applications in R}}, vol.~50.
\newblock Cambridge University Press.

\bibitem[{Bryc et~al.(2006)Bryc, Dembo \& Jiang}]{bryc2006hankel}
\textsc{Bryc, W.}, \textsc{Dembo, A.} \& \textsc{Jiang, T.} (2006).
\newblock {Spectral measure of large random Hankel, Markov and Toeplitz
  matrices}.
\newblock \textit{Annals of Probability} \textbf{34}, 1--38.

\bibitem[{Butucea \& Ingster(2013)}]{butucea2013detection}
\textsc{Butucea, C.} \& \textsc{Ingster, Y.~I.} (2013).
\newblock Detection of a sparse submatrix of a high-dimensional noisy matrix.
\newblock \textit{Bernoulli} \textbf{19}, 2652--2688.

\bibitem[{Cai et~al.(2017)Cai, Liang \& Rakhlin}]{cai2017computational}
\textsc{Cai, T.~T.}, \textsc{Liang, T.} \& \textsc{Rakhlin, A.} (2017).
\newblock Computational and statistical boundaries for submatrix localization
  in a large noisy matrix.
\newblock \textit{Annals of Statistics} \textbf{45}, 1403--1430.

\bibitem[{Cape et~al.(2024)Cape, Yu \& Liao}]{cape2024robust}
\textsc{Cape, J.}, \textsc{Yu, X.} \& \textsc{Liao, J.~Z.} (2024).
\newblock Robust spectral clustering with rank statistics.
\newblock \textit{Journal of Machine Learning Research} \textbf{25}, 1--81.

\bibitem[{Chatterjee(2006)}]{chatterjee2006generalization}
\textsc{Chatterjee, S.} (2006).
\newblock {A generalization of the Lindeberg principle}.
\newblock \textit{Annals of Probability} \textbf{34}, 2061--2076.

\bibitem[{Che(2017)}]{che2017universality}
\textsc{Che, Z.} (2017).
\newblock {Universality of random matrices with correlated entries}.
\newblock \textit{Electronic Journal of Probability} \textbf{22}, 1--38.

\bibitem[{Chen et~al.(2021)Chen, Chi, Fan \& Ma}]{chen2021spectral}
\textsc{Chen, Y.}, \textsc{Chi, Y.}, \textsc{Fan, J.} \& \textsc{Ma, C.}
  (2021).
\newblock Spectral methods for data science: a statistical perspective.
\newblock \textit{Foundations and Trends{\textregistered} in Machine Learning}
  \textbf{14}, 1--246.

\bibitem[{Chung \& Lee(2019)}]{chung2019weak}
\textsc{Chung, H.~W.} \& \textsc{Lee, J.~O.} (2019).
\newblock {Weak detection of signal in the spiked Wigner model}.
\newblock In \textit{Proceedings of the 36th International Conference on
  Machine Learning}, vol.~97. PMLR.

\bibitem[{Erd\H{o}s \& R\'{e}nyi(1959)}]{erdhos1959random}
\textsc{Erd\H{o}s, P.} \& \textsc{R\'{e}nyi, A.} (1959).
\newblock {On random graphs I}.
\newblock \textit{Publicationes Mathematicae, Debrecen} \textbf{6}, 290--297.

\bibitem[{Fan et~al.(2022)Fan, Fan, Han \& Lv}]{fan2022asymptotic}
\textsc{Fan, J.}, \textsc{Fan, Y.}, \textsc{Han, X.} \& \textsc{Lv, J.} (2022).
\newblock Asymptotic theory of eigenvectors for random matrices with diverging
  spikes.
\newblock \textit{Journal of the American Statistical Association}
  \textbf{117}, 996--1009.

\bibitem[{Fleermann(2019)}]{fleermann2019global}
\textsc{Fleermann, M.} (2019).
\newblock Global and local semicircle laws for random matrices with correlated
  entries.
\newblock \textit{Technical Report: see
  \text{https://ub-deposit.fernuni-hagen.de/}} .

\bibitem[{Fleermann \& Kirsch(2022)}]{fleermann2022central}
\textsc{Fleermann, M.} \& \textsc{Kirsch, W.} (2022).
\newblock The central limit theorem for weakly dependent random variables by
  the moment method.
\newblock \textit{arXiv preprint arXiv:2202.04717} .

\bibitem[{Fleermann et~al.(2021)Fleermann, Kirsch \&
  Kriecherbauer}]{fleermann2021almost}
\textsc{Fleermann, M.}, \textsc{Kirsch, W.} \& \textsc{Kriecherbauer, T.}
  (2021).
\newblock The almost sure semicircle law for random band matrices with
  dependent entries.
\newblock \textit{Stochastic Processes and their Applications} \textbf{131},
  172--200.

\bibitem[{F{\"u}redi \& Koml{\'o}s(1981)}]{furedi1981eigenvalues}
\textsc{F{\"u}redi, Z.} \& \textsc{Koml{\'o}s, J.} (1981).
\newblock The eigenvalues of random symmetric matrices.
\newblock \textit{Combinatorica} \textbf{1}, 233--241.

\bibitem[{G\H{o}tze et~al.(2015)G\H{o}tze, Naumov \&
  Tikhomirov}]{gotze2015limit}
\textsc{G\H{o}tze, F.}, \textsc{Naumov, A.} \& \textsc{Tikhomirov, A.} (2015).
\newblock Limit theorems for two classes of random matrices with dependent
  entries.
\newblock \textit{Theory of Probability \& Its Applications} \textbf{59},
  23--39.

\bibitem[{H{\'a}jek et~al.(1999)H{\'a}jek, {S}id{\'a}k \&
  Sen}]{hajek1999theory}
\textsc{H{\'a}jek, J.}, \textsc{{S}id{\'a}k, Z.} \& \textsc{Sen, P.~K.} (1999).
\newblock \textit{Theory of rank tests (second edition)}.
\newblock Academic Press.

\bibitem[{Holland et~al.(1983)Holland, Laskey \&
  Leinhardt}]{holland1983stochastic}
\textsc{Holland, P.~W.}, \textsc{Laskey, K.~B.} \& \textsc{Leinhardt, S.}
  (1983).
\newblock Stochastic blockmodels: First steps.
\newblock \textit{Social Networks} \textbf{5}, 109--137.

\bibitem[{Johnstone \& Paul(2018)}]{johnstone2018pca}
\textsc{Johnstone, I.~M.} \& \textsc{Paul, D.} (2018).
\newblock {PCA in high dimensions: an orientation}.
\newblock \textit{Proceedings of the IEEE} \textbf{106}, 1277--1292.

\bibitem[{Lei(2016)}]{lei2016goodness}
\textsc{Lei, J.} (2016).
\newblock {A goodness-of-fit test for stochastic block models}.
\newblock \textit{Annals of Statistics} \textbf{44}, 401--424.

\bibitem[{Leung \& Drton(2018)}]{leung2018independence}
\textsc{Leung, D.} \& \textsc{Drton, M.} (2018).
\newblock {Testing independence in high dimensions with sums of rank
  correlations}.
\newblock \textit{Annals of Statistics} \textbf{46}, 280--307.

\bibitem[{Li et~al.(2021)Li, Wang \& Li}]{li2021central}
\textsc{Li, Z.}, \textsc{Wang, Q.} \& \textsc{Li, R.} (2021).
\newblock {Central limit theorem for linear spectral statistics of large
  dimensional Kendall’s rank correlation matrices and its applications}.
\newblock \textit{Annals of Statistics} \textbf{49}, 1569--1593.

\bibitem[{Ma \& Wu(2015)}]{ma2015computational}
\textsc{Ma, Z.} \& \textsc{Wu, Y.} (2015).
\newblock Computational barriers in minimax submatrix detection.
\newblock \textit{Annals of Statistics} \textbf{43}, 1089--1116.

\bibitem[{Mossel et~al.(2015)Mossel, Neeman \& Sly}]{mossel2015reconstruction}
\textsc{Mossel, E.}, \textsc{Neeman, J.} \& \textsc{Sly, A.} (2015).
\newblock Reconstruction and estimation in the planted partition model.
\newblock \textit{Probability Theory and Related Fields} \textbf{162},
  431--461.

\bibitem[{Paul(2007)}]{paul2007asymptotics}
\textsc{Paul, D.} (2007).
\newblock Asymptotics of sample eigenstructure for a large dimensional spiked
  covariance model.
\newblock \textit{Statistica Sinica} \textbf{17}, 1617--1642.

\bibitem[{Perry et~al.(2018)Perry, Wein, Bandeira \&
  Moitra}]{perry2018optimality}
\textsc{Perry, A.}, \textsc{Wein, A.~S.}, \textsc{Bandeira, A.~S.} \&
  \textsc{Moitra, A.} (2018).
\newblock {Optimality and sub-optimality of PCA I: spiked random matrix
  models}.
\newblock \textit{Annals of Statistics} \textbf{46}, 2416--2451.

\bibitem[{Savage(1962)}]{savage1962bibliography}
\textsc{Savage, I.~R.} (1962).
\newblock \textit{Bibliography of Nonparametric Statistics}.
\newblock Harvard University Press.

\bibitem[{Shabalin et~al.(2009)Shabalin, Weigman, Perou \&
  Nobel}]{shabalin2009finding}
\textsc{Shabalin, A.~A.}, \textsc{Weigman, V.~J.}, \textsc{Perou, C.~M.} \&
  \textsc{Nobel, A.~B.} (2009).
\newblock Finding large average submatrices in high dimensional data.
\newblock \textit{Annals of Applied Statistics} \textbf{3}, 985--1012.

\bibitem[{Soshnikov(2004)}]{soshnikov2004poisson}
\textsc{Soshnikov, A.} (2004).
\newblock {Poisson statistics for the largest eigenvalues of Wigner random
  matrices with heavy tails}.
\newblock \textit{Electronic Communications in Probability} \textbf{9}, 82--91.

\bibitem[{Verzelen \& Arias-Castro(2015)}]{verzelen2015community}
\textsc{Verzelen, N.} \& \textsc{Arias-Castro, E.} (2015).
\newblock {Community detection in sparse random networks}.
\newblock \textit{Annals of Applied Probability} \textbf{25}, 3465--3510.

\bibitem[{Wainwright(2019)}]{wainwright2019high}
\textsc{Wainwright, M.~J.} (2019).
\newblock \textit{{High-dimensional statistics: a non-asymptotic viewpoint}},
  vol.~48.
\newblock Cambridge University Press.

\bibitem[{Wigner(1958)}]{wigner1958distribution}
\textsc{Wigner, E.~P.} (1958).
\newblock On the distribution of the roots of certain symmetric matrices.
\newblock \textit{Annals of Mathematics} \textbf{67}, 325--327.

\bibitem[{Wigner(1967)}]{wigner1967random}
\textsc{Wigner, E.~P.} (1967).
\newblock Random matrices in physics.
\newblock \textit{SIAM Review} \textbf{9}, 1--23.

\bibitem[{Wilcoxon(1945)}]{wilcoxon1945individual}
\textsc{Wilcoxon, F.} (1945).
\newblock Individual comparisons by ranking methods.
\newblock \textit{Biometrics Bulletin} \textbf{1}, 80--83.

\bibitem[{Wilcoxon(1946)}]{wilcoxon1946group}
\textsc{Wilcoxon, F.} (1946).
\newblock Individual comparisons of grouped data by ranking methods.
\newblock \textit{Journal of Economic Entomology} \textbf{39}, 269--270.

\bibitem[{Wilcoxon(1947)}]{wilcoxon1947probability}
\textsc{Wilcoxon, F.} (1947).
\newblock Probability tables for individual comparisons by ranking methods.
\newblock \textit{Biometrics} \textbf{3}, 119--122.

\bibitem[{Wu \& Wang(2022)}]{wu2022limiting}
\textsc{Wu, Z.} \& \textsc{Wang, C.} (2022).
\newblock {Limiting spectral distribution of large dimensional Spearman’s
  rank correlation matrices}.
\newblock \textit{Journal of Multivariate Analysis} \textbf{191}, 105011.

\bibitem[{Yuan et~al.(2022)Yuan, Yang \& Shang}]{yuan2022hypothesis}
\textsc{Yuan, M.}, \textsc{Yang, F.} \& \textsc{Shang, Z.} (2022).
\newblock Hypothesis testing in sparse weighted stochastic block model.
\newblock \textit{Statistical Papers} \textbf{63}, 1051--1073.

\bibitem[{Zhou(2007)}]{zhou2007asymptotic}
\textsc{Zhou, W.} (2007).
\newblock Asymptotic distribution of the largest off-diagonal entry of
  correlation matrices.
\newblock \textit{Transactions of the American Mathematical Society}
  \textbf{359}, 5345--5363.

\end{thebibliography}

\end{document}